\documentclass[a4paper,11pt]{article}
\pdfoutput=1
\usepackage{jheppub}
\usepackage{graphicx}
\usepackage{subfigure}
\usepackage{comment}
\usepackage{multirow}

\newcommand{\beq}{\begin{equation}}
\newcommand{\eeq}{\end{equation}}
\newcommand{\bea}{\begin{eqnarray}}
\newcommand{\eea}{\end{eqnarray}}

\def\tev{\,{\rm TeV}}
\def\gev{\,{\rm GeV}}

\def\fbi{\,{\rm fb}^{-1}}
\def\abi{\,{\rm ab}^{-1}}
\newcommand{\zp}{\ensuremath{Z'}}

\newcommand{\gzp}{\ensuremath{\Gamma_{Z'}}}

\newcommand{\pol}{\ensuremath{\mathcal{P}}}
\newcommand{\zpr}{\ensuremath{Z'}}
\newcommand{\mzp}{\ensuremath{M_{Z'}}}

\newcommand{\uprm}{\ensuremath{U(1)'}}

\newcommand{\ra}{\ensuremath{\rightarrow}}
\newcommand{\x}{\ensuremath{\times}}

\newcommand{\eeql}[1]{\label{#1}\eeq}
\newcommand{\refl}[1]{(\ref{#1})}

\title{Diagnosis of a New Neutral Gauge Boson at the LHC and ILC for Snowmass 2013\\
}

\author[a]{Tao Han}
\author[b,c]{, Paul Langacker}
\author[a]{, Zhen Liu}
\author[d]{and Lian-Tao Wang}

\affiliation[a]{Pittsburgh Particle physics, Astrophysics, and Cosmology Center, \\
Department of Physics and Astronomy, University of Pittsburgh, \\
3941 O'Hara St., Pittsburgh, PA 15260, USA}
\affiliation[b]{School of Natural Science, Institute for Advanced Study, \\
Einstein Drive, Princeton, NJ 08540, USA}
\affiliation[c]{Department of Physics, Princeton University, \\
Princeton, NJ 08544, USA}
\affiliation[d]{Department of Physics, Enrico Fermi Institute, \\
and Kavli Institute for Cosmological Physics,\\
University of Chicago, Chicago, IL 60637-1434, USA}

\abstract{A \uprm\ or \zpr\  is  generic in many scenarios of physics beyond the Standard Model, such as string theory compactifications, GUTs, extra-dimensions, compositeness, dynamical electroweak symmetry breaking, dark-sector models, etc. We study the potential of probing a TeV-scale \zp\ with electroweak couplings in future experiments. In particular, we focus on two  scenarios: (1) If a $\zp$ is discovered at the LHC, what is the potential of measuring its mass and width and to distinguish between benchmark models utilizing various observables, especially asymmetries, at a high luminosity LHC and the ILC. (2) If the $\zp$ is not accessible as a clear resonance signal, what is the exclusion reach at the ILC.
}

\keywords{$\zp$, LHC, ILC}

\begin{document}
\maketitle

\newpage

\section{\boldmath{\zpr} Parameters}
\label{sec:theo}

Additional colorless vector gauge  bosons (\zpr) occur in many extensions of the Standard Model (SM),
in part because it is generically harder to break additional abelian $U(1)^\prime$ factors than  non-abelian ones\footnote{For  reviews, see~\cite{Langacker:2008yv,Cvetic:1995zs,Hewett:1988xc,Leike:1998wr}.  Specific properties are reviewed in~\cite{Cvetic:1997wu,Erler:2009jh,Langacker:2009im,Nath:2010zj,Jaeckel:2010ni,delAguila:2010mx,Diener:2011jt}.}.
The existence of a \zpr\ could have many other possible implications,
including an NMSSM-like solution to the $\mu$ problem (and the possibility of electroweak baryogenesis), new $F$ and $D$ term contributions to the lightest scalar mass, an additional Higgs singlet,
additional neutralinos (with collider and dark matter consequences), new vector (under the SM) fermions for anomaly cancellation, and many possibilities for neutrino mass.
Other possibilities involve interactions with dark matter, the mediation of supersymmetry
breaking, FCNC (for family non-universal couplings), associated charged $W'$ s,
and the production of superpartners and exotics.
The \zpr\ couplings could also give clues about a possible embedding of the \uprm\ into a more fundamental underlying theory.
Although \zpr s can occur at any scale and with couplings ranging from extremely weak to strong, we concentrate here on TeV-scale masses with couplings not too different from electroweak,  which might therefore  be observable at the LHC or future colliders.

Following the notation in~\cite{Langacker:2008yv}, we define the couplings of the SM and additional neutral gauge bosons to fermions by
\beq
-L_{NC}=eJ^\mu_{em}A_\mu+g_1J_1^\mu Z^0_{1\mu} + g_2 J^\mu_2 Z_{2 \mu}^0,
\eeql{lag}
with
\beq
J_\alpha^\mu=\sum\limits_i \bar f_i \gamma^\mu [\epsilon_L^{\alpha i}P_L + \epsilon_R^{\alpha i} P_R] f_i.
\eeql{current}
The SM ($Z^0_1$) parameters are $g_1= g/\cos\theta_W$ and $\epsilon_{L}^{1i} =t^i_{3L}-\sin^2 \theta_W q^i$,  $\epsilon_{R}^{1i} =-\sin^2 \theta_W q^i$,
where  $q^i$ is the electric charge of $f_i$ in units of $|e|$ and $t^i_{3L}=\pm 1/2$ is the third component of weak isospin.
We will absorb $g_\alpha$ into the chiral charges\footnote{The gauge coupling $g_2$ is not really a separate parameter, because it can be absorbed in the chiral couplings, as in
\refl{charges}. However, the separate extraction of $g_2$ would become meaningful if the charges were established to correspond to an embedding in a nonabelian group of some other model with well-defined normalization, such as the $E_6$ and LR models.} by defining
\beq g_{L,R}^{1i} \equiv g_1 \epsilon_{L,R}^{1i}, \qquad    g_{L,R}^{2 i} \equiv g_2 \epsilon_{L,R}^{2 i}. \eeql{charges}
When it does not cause confusion we will drop the superscript $2$ on  $g_{L,R}^{2 i} $.
It will also be convenient to define the vector and axial couplings and the asymmetry parameters
\beq
g^i_{V,A} \equiv g^i_L \pm g^i_R,\qquad   A_i \equiv  \frac{g_L^{i\,2} - g_R^{i\,2} }{g_L^{i\,2}  +g_R^{i\,2} }
= \frac{2\, g_V^i g_A^i}{g_V^{i\, 2} + g_A^{i\, 2}},\eeql{vacoup}
for $i=u, d, e, \nu, \cdots$.
Analogous definitions hold for the $g_{L,R}^{1i}$.

Assuming negligible  (ordinary and kinetic)  $Z-Z^\prime$ mixing~\cite{Erler:2009jh,delAguila:2010mx,Diener:2011jt}) and family universality, the relevant
\zpr\ parameters are \mzp, \gzp, and the chiral couplings $g^i_{L,R}$ for $i=u, d, e,$ and $\nu$.
A lower bound on \gzp\ (the ``minimal'' width) can be calculated in terms of the other parameters from the decays into the SM fermions,
but a larger \gzp\ is possible due to decays into Higgs particles, superpartners, right-handed neutrinos, exotic fermions (such as those needed in some \zpr\ models for anomaly cancellation), or other Beyond the Standard Model (BSM) particles~\cite{Kang:2004bz,Chang:2011be}.
We will usually assume as well that the \uprm\ charges commute\footnote{One exception is the benchmark sequential model, in which  $g_{L,R}^{2 i}=g_{L,R}^{1 i}$. This could possibly emerge from a diagonal embedding of the SM in a larger group, or for Kaluza-Klein excitations in an extra-dimensional theory.} with $SU(2)$, so that there are only five
relevant chiral charges,
\beq
g_L^u=g_L^d\equiv g_L^q, \quad g_R^u,\quad g_R^d, \quad g_L^e=g_L^\nu\equiv g_L^\ell, \quad g_R^e. \eeql{funiv}

Ideally, one would like to determine these, as well as \mzp\ and \gzp, in a model-independent way
from collider as well as existing and future precision data. In practice, the existing limits are sufficiently stringent
that we may have to resort to considering specific benchmark models.
For illustration, we will consider the well-known $\chi$, $\psi$, and LR models, associated with the
breakings  $SO(10)\ra SU(5)\x U(1)_\chi$, $E_6 \ra SO(10)\x U(1)_\psi$,
and $SU(2)_L \x SU(2)_R \x U(1)_{B-L} \ra SU(2) \x U(1)_Y \x U(1)_{LR}$ (for $g_R=g$), respectively. We will also consider
 $Z_\eta= \sqrt{\frac{3}{8}}Z_\chi-  \sqrt{\frac{5}{8}}Z_\psi$, associated with a certain compactification of the heterotic string,
 and the B-L model\footnote{The $B-L$ charge usually occurs in a linear combination with $T_{3R}=Y-\frac{B-L}{2}$, where $Y=Q-T_{3L}$, as in the $\chi$ and LR models. Here we consider a simple $B-L$ charge as an example of a purely vector coupling.} with charge $(B-L)/2$. The charges for these benchmark models are listed in Table~\ref{tab:benchmark}.
 For the $E_6$, LR, and B-L models we will take for the reference value of  $g_2$ the GUT-normalized hypercharge coupling
 \beq g_2 = \sqrt{\frac{5}{3}}\, g \tan\theta_W \sim 0.46, \eeql{refg2}
 which is an approximation to the simplest $E_6$ prediction~\cite{Robinett:1982tq} for the GUT models and follows
 for $g_R=g$ in $SU(2)_L \x SU(2)_R \x U(1)_{B-L}$. We will also consider the sequential model with $g_2=g_1$ and $\epsilon^{2 i}_{LR} = \epsilon^{1 i}_{LR}.$

\begin{table}[htb]
  \centering
    \begin{tabular}{|c|c|c|c|c|c|c|c|}
    \hline
          & $\chi$ & $\psi$ & $\eta$ & LR & B-L & \multicolumn{2}{|c|}{SSM}\\
    \hline
    $D$ & $2\sqrt{10}$ & $2\sqrt{6}$ & $2\sqrt{15}$ & $\sqrt{5/3}$ & 1 & \multicolumn{2}{|c|}{1} \\ \hline
 \multirow{2}{*}{$\hat \epsilon_L^q$} & \multirow{2}{*}{--1}    & \multirow{2}{*}{1}     & \multirow{2}{*}{--2}    & \multirow{2}{*}{--0.109} & \multirow{4}{*}{1/6} & $\hat \epsilon_L^u$ & $\frac 1 2 -\frac 2 3{\sin}^2\theta_W$\\ \cline{7-8}
         &   &   &   &   &   & $\hat \epsilon_L^d$ & $-\frac 1 2 +\frac 1 3{\sin}^2\theta_W$\\ \cline{1-5} \cline{7-8}
    $\hat \epsilon_R^u$ & 1     & --1    & 2     & 0.656  &  & $\hat \epsilon_R^u$ & $- \frac 2 3 {\sin}^2\theta_W$\\ \cline{1-5} \cline{7-8}
    $\hat \epsilon_R^d$ & --3    & --1    & --1    & --0.874 &  & $\hat \epsilon_R^d$ & $ \frac 1 3 {\sin}^2\theta_W$\\ \hline
    \multirow{2}{*}{$\hat \epsilon_L^l$} & \multirow{2}{*}{3}     & \multirow{2}{*}{1}     & \multirow{2}{*}{1}     & \multirow{2}{*}{0.327}  & \multirow{3}{*}{--1/2} & $\hat \epsilon_L^\nu$ & $\frac 1 2 $\\ \cline{7-8}
           &   &   &   &   &   & $\hat \epsilon_L^e$ & $-\frac 1 2 + {\sin}^2\theta_W$\\ \cline{1-5} \cline{7-8}
    $\hat \epsilon_R^e$ & 1     & --1    & 2     & --0.438 &  & $\hat \epsilon_R^e$ & $ {\sin}^2\theta_W$\\ \hline
    \end{tabular}%
  \label{tab:benchmark}%
\caption{Benchmark models and couplings, with $\epsilon^i_{L,R} \equiv \hat \epsilon^i_{L,R} /D$.}
\end{table}%

\section{Case 1: \boldmath{\zpr} Observable at the LHC} \label{sec:case1}
Typical \zpr\ models with electroweak couplings should be observable\footnote{The reach is reduced if the dilepton branching ratios
are significantly reduced due to BSM decay channels~\cite{Kang:2004bz,Chang:2011be}.} at the LHC as resonances
in the dilepton channels for masses up to $\sim$4-5 TeV for $\sqrt{s} = 14$ TeV and an integrated luminosity of 100 fb$^{-1}$. There have been extensive studies of diagnostic possibilities\footnote{See, for example,~\cite{Langacker:1984dc,Czarnecki:1990pe,delAguila:1993ym,DelAguila:1993rw,DelAguila:1995fa,Dittmar:2003ir,Kang:2004bz,Carena:2004xs,Weiglein:2004hn,Petriello:2008zr,Godfrey:2008vf,Osland:2009tn,Li:2009xh,Diener:2009vq,Diener:2010sy,Gopalakrishna:2010xm,Chang:2011be,Erler:2011ud,Berger:2011hn,Chiang:2011kq,Accomando:2013sfa,Berger:2013fya}. Other studies are reviewed in~\cite{Langacker:2008yv,Leike:1998wr,Nath:2010zj}.} of the $Z'$ couplings at the LHC utilizing the cross sections
\beq \sigma^{f } \equiv \sigma[f \bar f] \equiv  \sigma_{pp\to\zp\to f \bar f} = \sigma_{\zpr} B(\zpr\ra f \bar f)
\eeql{csec}
for decays into the final state $f \bar f$
for $f=\ell, \tau, t, b$ (with $\ell=e, \mu$), as well as forward-backward  or charge asymmetries, rapidity distributions, and possible final state polarizations  for $\tau^- \tau^+$ or $t \bar t$. Other possible probes include
\gzp\ from the lineshape, and various rare decay modes and associated productions. It was generally concluded that significant diagnostic probes of the couplings would be possible for \zpr\ masses up to around 2.5 TeV.

However, ATLAS~\cite{ATLAS:2013-017} and CMS~\cite{CMS:EXO12061} have already excluded  dilepton resonances corresponding to standard
benchmark \zpr s below $\sim$2.5-2.9 TeV, so even if a \zpr\ is observed in future LHC running it will be
difficult to carry out detailed diagnostics. We have therefore re-examined what might be learned for a relatively heavy
\zpr, allowing for high integrated luminosities of 300 and 3000 fb$^{-1}$ at the LHC, in combination
with observations at the ILC with $\sqrt{s}=500~\gev$ and integrated luminosity of $500~\fbi$,
or at 1 TeV with 1000 fb$^{-1}$, for fixed $e^\mp$ polarizations. We also consider
the possibility of additional ILC running with reversed polarizations.
We consider two illustrative cases: (1) a 3 TeV \zpr\ observed directly at the LHC and indirectly at the ILC; (2) a more massive \zpr\ observed only by indirect effects at the ILC.
Future studies will also include indirect constraints from existing and future precision experiments.

\subsection{LHC Searches}\label{sec:LHCsearches}
The formalism relevant to the production and decay of a \zpr\ at the LHC is summarized in
Appendix \ref{app:LHC}. We assume in this section that a narrow colorless resonance has been observed as
a peak in the $\ell^- \ell^+$ distribution at the LHC at mass \mzp, and that the lepton angular distribution has identified that the resonance has spin-1~\cite{Dittmar:2003ir,Osland:2009tn}. Assuming family universal couplings and neglecting $Z-\zpr$
mixing (known to be small from precision electroweak studies~\cite{Erler:2009jh,delAguila:2010mx,Diener:2011jt}),
there remain to be determined the five chiral couplings in \refl{funiv} as well as \gzp. Ideally, one would like to determine these in as model-independent a way as possible.

The simplest observables (other than \mzp) are the cross sections
$\sigma^{f }  = \sigma_{\zpr} B(\zpr\ra f \bar f)$ after subtracting backgrounds, especially for $f=e, \mu$.
However, the  cross sections have uncertainties from the parton distribution functions (PDFs), higher-order terms,
and the luminosity. Furthermore, they are inversely proportional to \gzp, as in \refl{lhc14}, so they do not allow a determination of the absolute couplings, even in principle. Also, the leptonic rates depend only on a linear combination of the $u$ and $d$ couplings (roughly 2 to 1 at the LHC), unless there is significant information from the rapidity distribution (which is unlikely at the LHC).

The \gzp\ ambiguity can be eliminated and the PDF/higher order uncertainties
can be reduced by considering ratios of observables. If one can tag the $f=b$ and $t$ final states well enough\footnote{The total dijet rate may be impossible to observe for a \zpr\ with electroweak coupling.}
then the ratios of the rates for $f=\ell, b, t$ could in principle determine the ratios of
$g_L^{q\, 2}+ g_R^{u\, 2}, g_L^{q\, 2}+ g_R^{d\, 2},$ and $g_L^{\ell\, 2}+ g_R^{e\, 2}$ (again assuming family universality).
These could be promoted to absolute measurements if \gzp\ can be extracted from the lineshape, since
the product $\sigma^{f } \gzp =  \sigma_{\zpr} \Gamma(\zpr\ra f \bar f)$ depends only on the absolute couplings.

Forward-backward or charge asymmetries could yield additional information. From \refl{lhc20} we see that $g_R^{f\, 2}/g_L^{f\, 2}$ can be determined for $f=\ell, u, d$ if charge identification  is available for $\ell, t, b$, respectively.
This is again independent of \gzp\ and involves reduced PDF uncertainties. Final state polarization
effects for $f=\tau$ or $t$ could carry complementary information, which could  increase the accuracies of the determinations  and/or help to test our assumption of family universality.
Off-pole interference with standard model
(mainly $\gamma$ and $Z$) backgrounds could also in principle yield information such as the signs of the couplings~\cite{Erler:2011ud,Accomando:2013sfa}.

As stated previously, however, the existing LHC limits are sufficiently strong that it will most likely not be
possible to obtain significant model-independent determinations of the couplings from the LHC alone.
Nevertheless, some of the observables could at least allow discrimination between the benchmark models.

\subsubsection{Leptonic Final States}\label{sec:leptonic}

The leptonic final states are very clean at the LHC. The standard model dilepton background is at the attobarn level, negligible compared to the femtobarn-level signal. We tabulate the cross sections and total widths for our benchmark models in Table~\ref{tab:lept} for $\mzp=3$ TeV.  These widths are ``minimal''; if the $\zp$ can decay into final states other than standard model fermions, the total width will increase, resulting in a suppression of the standard model fermion branching fractions  as well as the appearance of new visible (invisible) final states like $W_R^+W_R^-$ (sterile $\nu_R^c\nu_R$).

\begin{table}[htb]
  \centering
    \begin{tabular}{|c|c|c|c|c|c|c|c|}
    \hline
          & $\chi$ & $\psi$ & $\eta$ & LR & B-L & SSM \\
    \hline
    width (3 TeV $\zp$) (GeV) & 34.7 & 15.7 & 18.9 & 61.4 & 27.4 & 88.7 \\ \hline
    $\sigma^e$ (fb) & 0.850 & 0.430 & 0.503 & 1.006 & 1.004 & 1.602 \\ \hline
    \end{tabular}%
\caption{The minimal widths for the benchmark $\zp$ models and the cross sections $\sigma^{e }  = \sigma [e^-e^+] =\sigma_{\zpr} B(\zpr\ra e^- e^+)$ at the ($14~\tev$) LHC for dielectron
final states in the mass window  $2.8-3.2~\tev$.
The acceptance of the electron-positron pair is taken to be $78\%$.
}
\label{tab:lept}
\end{table}%

\begin{figure}[htb]
\centering
\subfigure{
\includegraphics[bb=0.0in 0in 7in 6.0in,width=200pt]{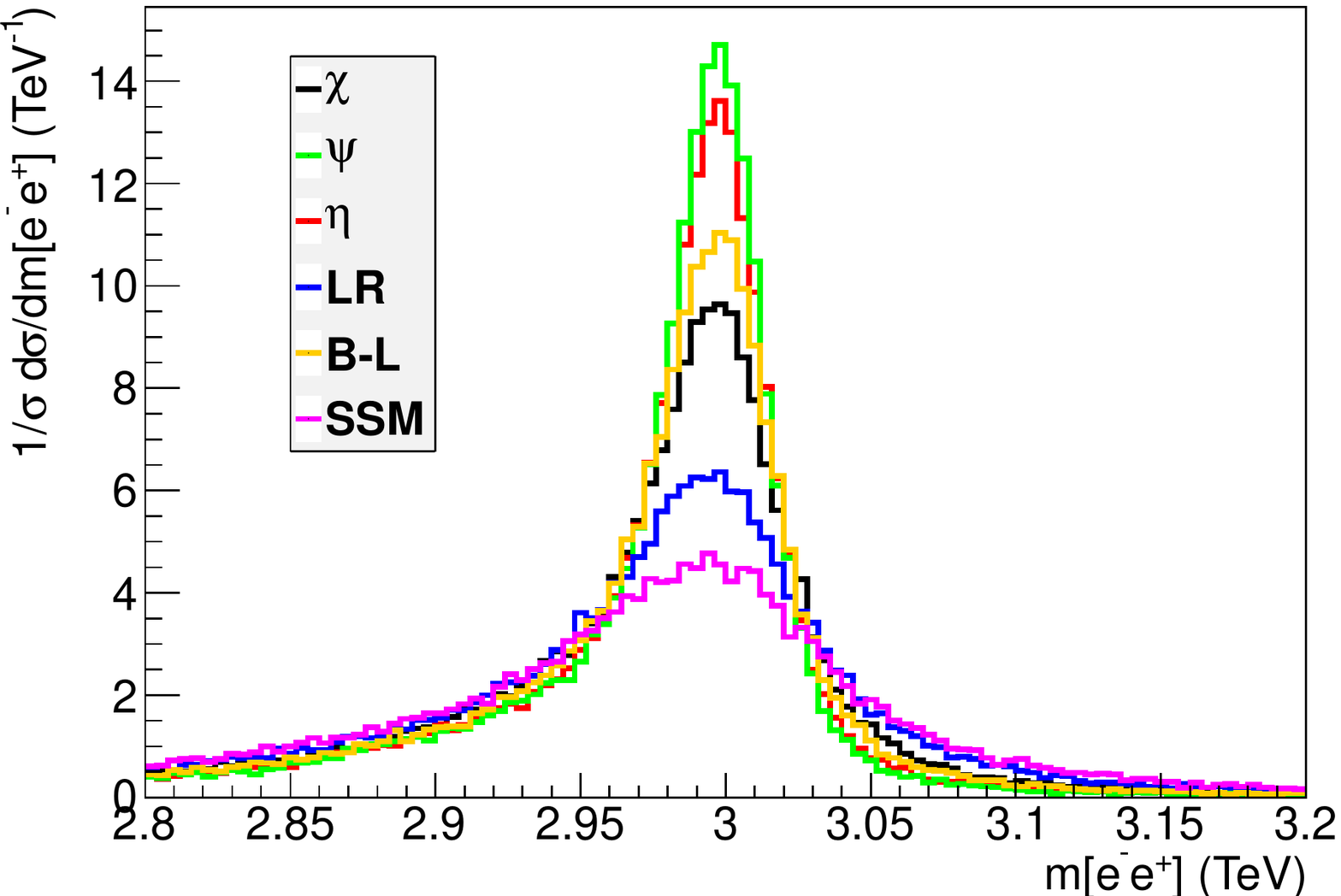}}
\subfigure{
\includegraphics[bb=0.0in 0in 7in 6.0in,width=200pt]{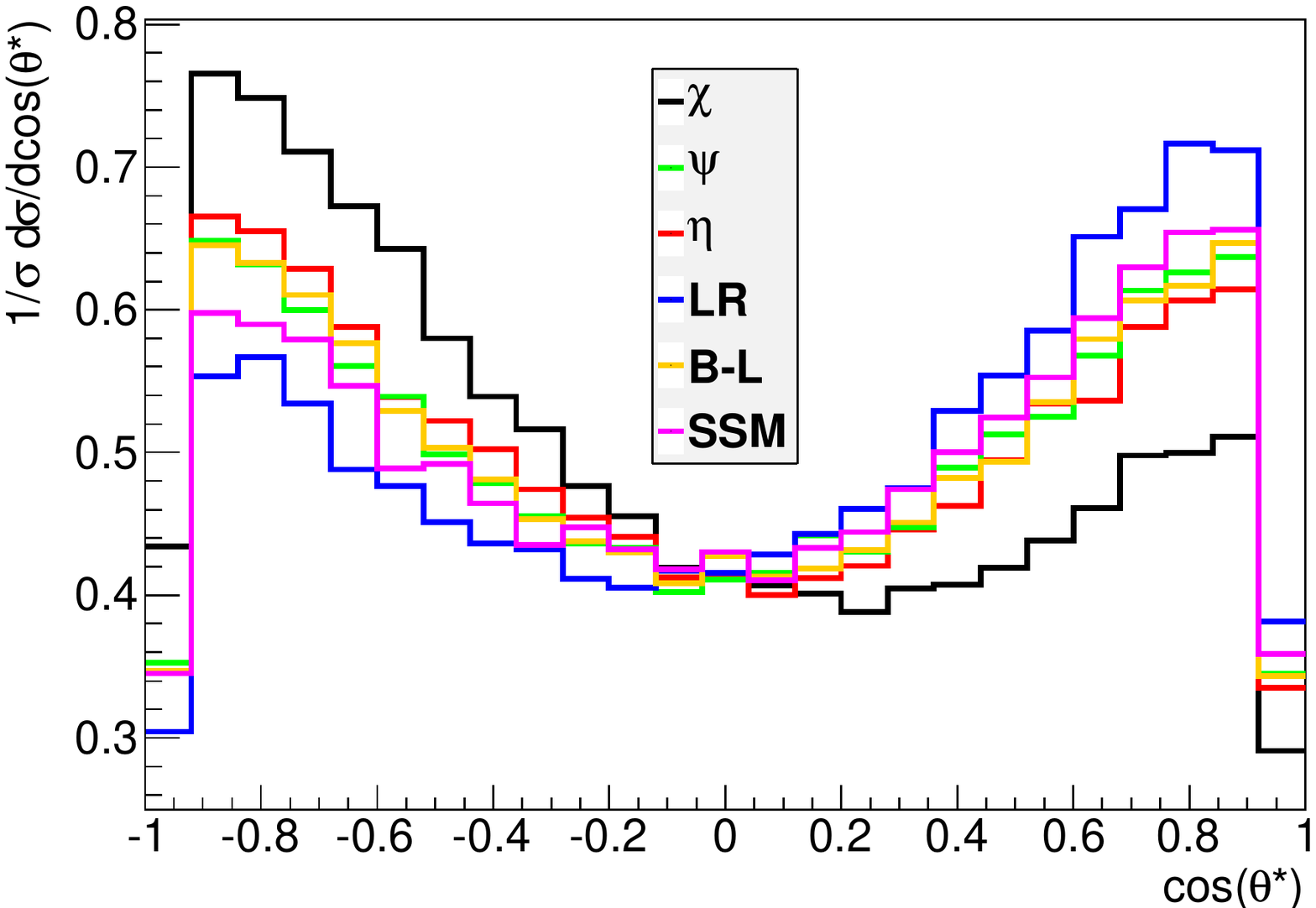}}
\caption[]{ {\it Left panel}: the invariant mass distribution of the dielectron system for the benchmark models for a  $3$ TeV  $\zp$  at  the LHC at $14$ TeV;
{\it Right panel}: the angular distribution of the electrons in the CM frame with respect to the rapidity (boost) direction of the system in the lab frame, integrated over the dielectron rapidity $y$.
}
\label{fig:LHCdimuon}
\end{figure}

We simulate the signal and background events using MadGraph5~\cite{Alwall:2011uj} with input model files generated by FeynRules~\cite{Christensen:2008py}, using proton parton distribution functions (PDF) set  CTEQ6l1~\cite{Pumplin:2002vw}. The generated events then pass through Pythia6~\cite{Sjostrand:2006za} to perform parton showers and then Delphes~\cite{Ovyn:2009tx} for detector simulation using the Snowmass Delphes3 card.
We show the invariant mass distributions and the angular distributions in the center of mass (CM) frame of the dielectron system for these benchmark models in Figure~\ref{fig:LHCdimuon}. One can extract the mass, width, and  total rate $\sigma^{e }  $ from the invariant mass distribution as shown in the left panel\footnote{The rapidity distribution of the dielectron pair  could in principle be useful for separating the effects of the $u$ and $d$. In practice, however, there is little sensitivity for $\mzp\gtrsim 3$ TeV.}.
The dimuon final state is similar. The energy resolution for high energy muons is worse than for electrons according to the Snowmass detector simulation. As a result, dimuon final states will provide additional statistics for   $\zp$ discovery but won't contribute much to the  mass and width determinations.

The forward-backward asymmetry $A_{FB}$, defined in \refl{lhc20} (which is equivalent to the charge asymmetry $A_c$ in \refl{lh21}),
can be obtained directly by counting, from the charge asymmetry, or by fitting to the angular distribution  shown in the right panel of Figure~\ref{fig:LHCdimuon} for the benchmark models.
From  \refl{lhc20} one sees that $A_{FB}$ is sensitive to the difference between the left and right- chiral couplings-squared of the leptons and of the quarks.
Of course, there is no forward-backward asymmetry in a $pp$ collider at zero \zpr\ rapidity $y$, but there can be an asymmetry for nonzero $y$.
We define the forward direction with respect to the rapidity (boost) direction of the \zpr\, or equivalently of the dielectron system. The (mainly valence)
quark direction is usually  the same as the boost direction at the LHC. However,  around $20\%$ of the events have the anti-quark direction along the boost direction (the contamination factor). This contamination factor varies for different PDF sets, adding additional theoretical uncertainties. It also varies somewhat with the \zpr\ model because of the different relative couplings of up-type and down-type quarks.

\begin{figure}
\subfigure{
\includegraphics[bb=0.0in 0in 7in 6.0in,width=200pt]{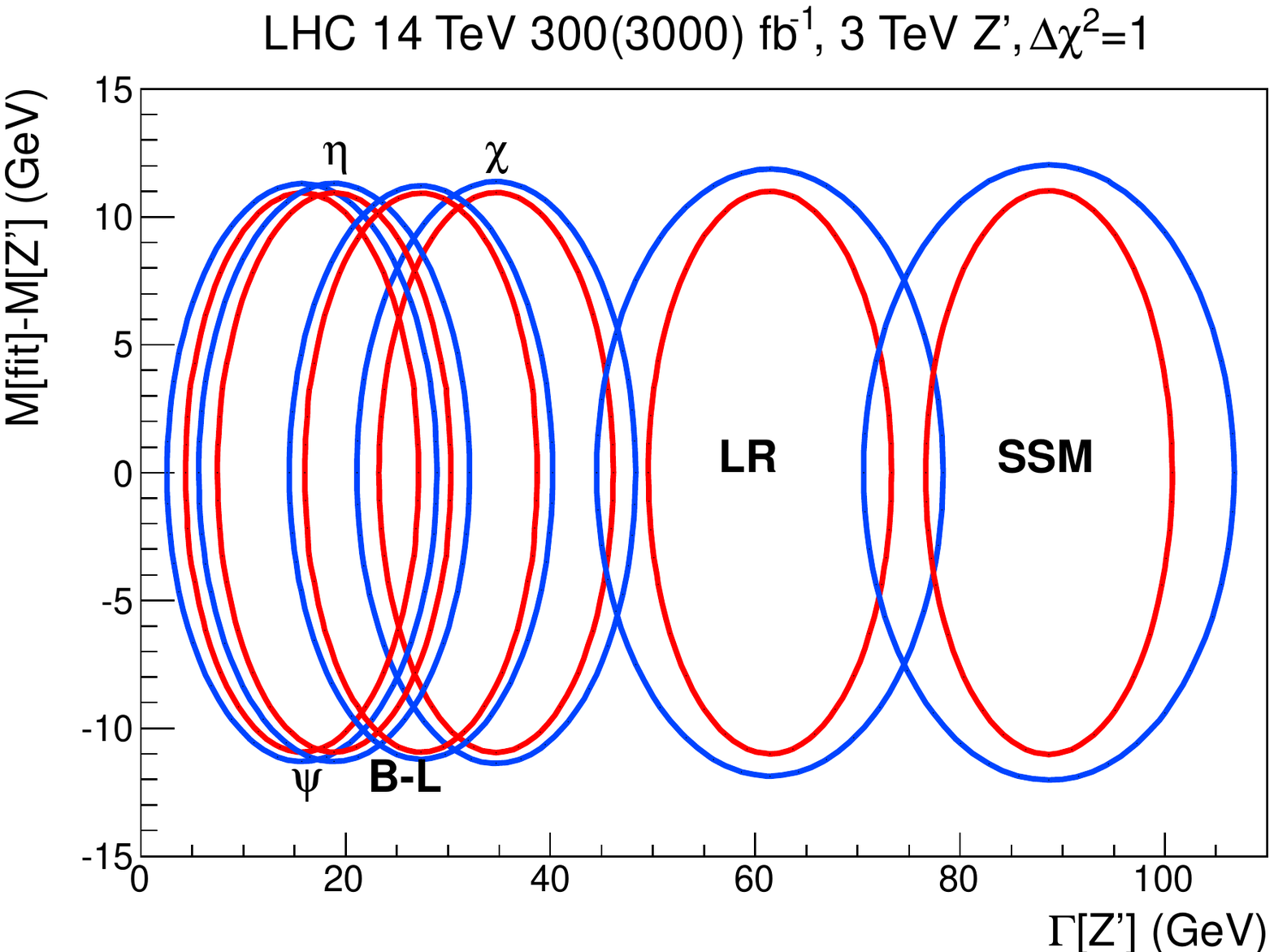}}
\subfigure{
\includegraphics[bb=0.0in 0in 7in 6.0in,width=200pt]{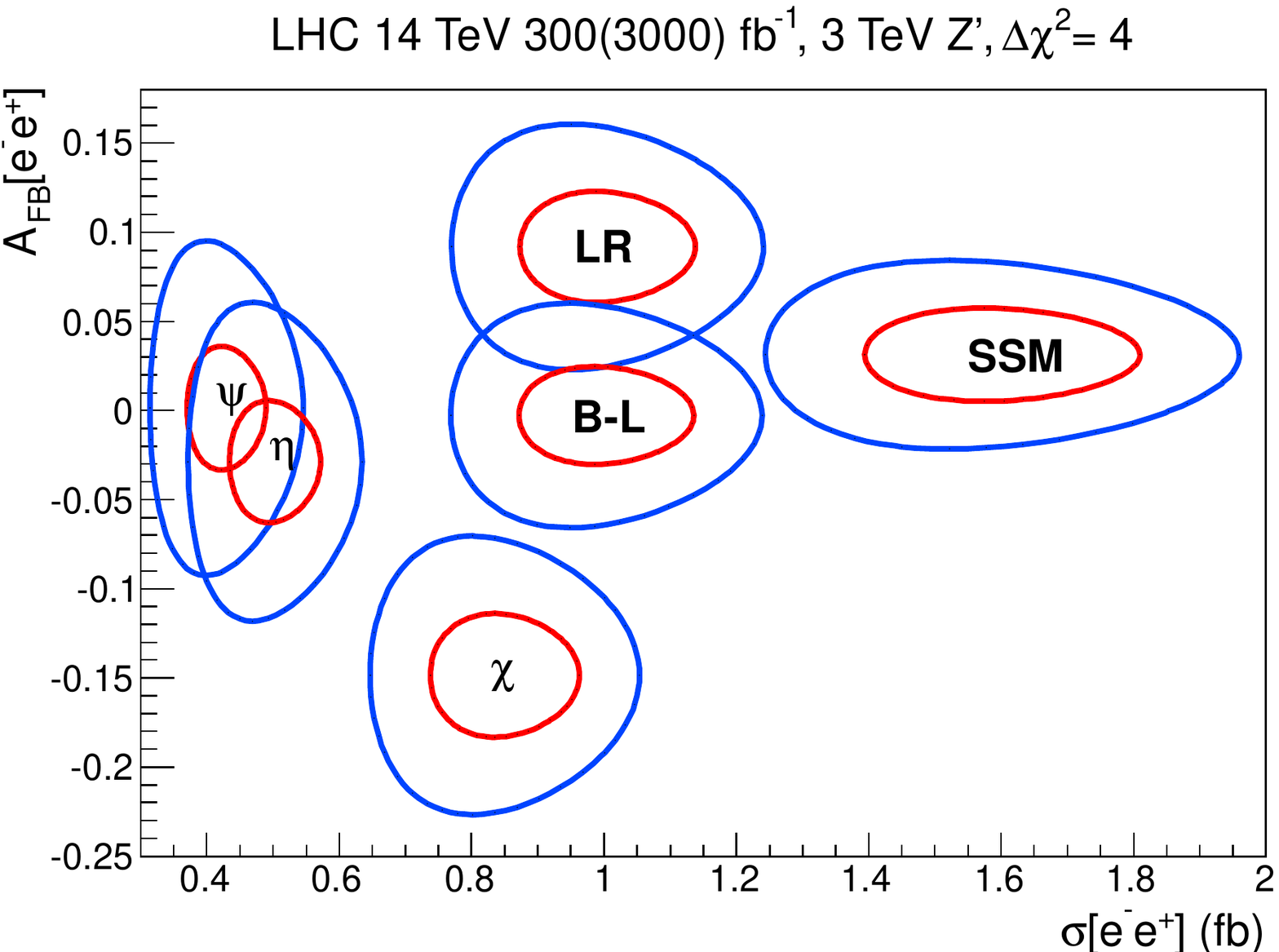}}
\caption[]{The results for $pp\to\zp\to e^-e^+$ with dielectron invariant mass from $2.8-3.2~\tev$. {\it Left panel}:  $\Delta \chi^2=1$ contours for the fitted width versus mass for the LHC at $300~\fbi$ and $3000~\fbi$.
{\it Right panel}: $\Delta\chi^2=4$ contours of the simulated forward-asymmetry versus the cross section.
}
\label{fig:LHCmmfit}
\end{figure}

In order to estimate the sensitivity to the \zpr\ parameters, we have simulated the lineshape and angular distributions for each of our benchmark models, assuming the minimal width, and then ``fit'' to the simulated data to determine the uncertainties in the extracted parameters.
We show the fitting results for the masses and widths of the $\zp$ in the { left panel} of Figure~\ref{fig:LHCmmfit}, and the simulated cross section and forward-backward asymmetries in the {right panel}. The two contours are for the  LHC at $14~\tev$ and $300\fbi$ (blue) and $3000\fbi$ (red). The fitting for the  mass and width is  model-independent. We fit the invariant mass distribution by a Breit-Wigner resonance convoluted with a Gaussian distribution for the smearing  from the  electron energy resolution. We assume $0.7\%$ systematic uncertainties for the  mass and width ($\sqrt 2$ times the electron energy resolution $0.5\%$).
We see that \mzp\ can be reproduced to around $10~\gev$, i.e., better than one percent. \gzp\ can also be determined to around $10~\gev$, but from Table~\ref{tab:lept}
and Figure~\ref{fig:LHCmmfit} this is very crude (e.g., 30-60\%) for the minimal widths in most of the benchmark models.
The total width and mass precision is dominated by systematic uncertainties: one can see that the improvement from $300~\fbi$ to $3000~\fbi$ is not significant.
Nevertheless, the LHC is the only planned facility that can measure these quantities to any precision\footnote{In principle the mass could be determined indirectly, e.g., by comparing results from the ILC at different energies. However, the ILC
  sensitivity is small for a multi-TeV $\zp$ mass.}.

We also show the forward-backward asymmetry  and cross section  determinations\footnote{The uncertainties in \gzp\ are too large to obtain useful model-independent constraints from $\sigma^e \, \gzp$.}.
In addition to the statistical uncertainties, we take the systematic uncertainties $10\%\oplus2\%~(6\%\oplus2\%)$ for the cross section for the LHC at $300~\fbi~(3000~\fbi)$. The $10\%~(6\%)$ are the correlated uncertainties (e.g., PDF  and luminosity uncertainties) that will cancel when taking the ratios of cross sections, leaving $2\%$ systematics for the forward-backward asymmetry. $A_{FB}$ can be determined very well for  asymmetric models such as the $Z_\chi$ and  $Z_{ LR}$, approximately $20\%$ ($5\%$) at the  LHC $14~\tev$ with $300~\fbi$ ($3000~\fbi$). The absolute error is comparable for the other (more symmetric) models.
The contours in Figure~\ref{fig:LHCmmfit} indicate that there is some reasonable possibility of distinguishing some of the benchmark models with minimal width
at the LHC 14 TeV. However, there is not much possibility for model-independent studies
based on the dielectron observables alone.

\subsubsection{Hadronic Final States}\label{sec:hadronic}

The hadronic final states of the $3$ TeV $\zp$ are particularly important. Once combined with the leptonic channels, under the assumption of family universality, one can in principle obtain the absolute values of the $\zp$ coupling strength to both leptons and hadrons. On the other hand, one faces the difficulties of huge QCD backgrounds. In this section we discuss the possibility of observing these channels at the LHC.

We  list the parton level cross section for both signal and irreducible background at the LHC $14~\tev$ in Table~\ref{tab:LHCqq}. The cross sections for these models for the dijet final state, including up, down, charm and strange quarks, are at the  femtobarn level. The QCD background, after preliminary cuts, is $\sim$1000 times larger than the signal. More strict cuts and selection criteria may help improve this channel, but nevertheless the dijet channel is not promising.

We are particularly interested in the third generation final states. Heavy quark tagging techniques make it possible to observe these channels. Moreover, they can determine the (family universal) $\zp$ couplings to up-type quarks and down-type quarks. In the case that top quark charge and/or polarization tagging is available, one would be able to obtain constraints on the chiral couplings of the $\zp$.
 On the other hand, the top quark signal is statistically very limited, as shown in the table. The top tagging and mis-tagging rates in this highly boosted scenario require further investigation. Thus we only list its parton level cross section and not discuss  backgrounds.

\begin{table}[htb]
  \centering
    \begin{tabular}{|c|c|c|c|c|c|c|}
    \hline
          & $\chi$ & $\psi$ & $\eta$ & LR & B-L & SSM \\
    \hline
    $\sigma_{2j}^{SM}$ (fb) & \multicolumn{6}{c|}{$1.4\times10^6$} \\ \hline
    $\sigma_{2j;{\rm cut}}^{SM}$(fb) & \multicolumn{6}{c|}{$5.1\times10^3$} \\ \hline
    $\sigma_{2j}^{\zp}$ (fb) & 6.0 & 5.6 & 8.3 & 21 & 1.4 & 19 \\ \hline \hline
    $\sigma_{2b}^{\zp}$ (fb) & 2.9 & 1.6 & 1.9 & 7.8 & 0.4 & 6.2 \\ \hline
    $\sigma_{2b}^{SM+\zp}$ (fb) & 5.5 & 3.7 & 3.9 & 10 & 2.3 & 8.7 \\ \hline \hline
    $\sigma_{2t}^{\zp}$ (fb) & 0.7 & 1.7 & 3.2 & 5.8 & 0.5 & 7.0 \\ \hline
    \end{tabular}%
\caption{Parton level cross sections at the LHC $14~\tev$. We only select events with final state dijet and bottom pair invariant mass in the  window $2.9-3.1~\tev$.
$\sigma^{SM}_{2j;{\rm cut}}$ are with cuts $h_t$ (scalar sum of jets' $p_T$s) $>500~\gev,\ p_T>200~\gev,$ and $y_j<2$.
}
\label{tab:LHCqq}%
\end{table}%

\begin{table}[htbp]
  \centering
    \begin{tabular}{|c|c|c|c|c|c|c|c|c|}
    \hline
          & QCD Dijet & SM $b\bar b$ & $\chi$   & $\psi$   & $\eta$   & LR & B-L & SSM \\ \hline
    $\sigma$ (fb)    & 36300 & 12.1 & 3.44 & 1.73 & 2.03 & 10.8 & 0.45 & 9.74 \\ \hline
    $\epsilon_{b}$ (\%)& 0.561 & 27.6 & 30.7 & 30.1 & 30.2 & 29.7 & 30.7 & 28.7 \\ \hline
    $\epsilon_{P^{b}_t}$ (\%)  & 0.0365 & 6.80 & 9.07 & 9.14 & 9.34 & 8.15 & 9.56 & 7.63 \\ \hline
\hline
    $\sigma_{\rm eff}$ (fb) & 11.78 & 0.82  & 0.31 & 0.16 & 0.19 & 0.88 & 0.04 & 0.74 \\ \hline
    $\frac S {\sqrt B} @ $ $0.3~\abi$ &       &       & 1.5  & 0.8  & 0.9  & 4.3 & 0.2 & 3.6 \\ \hline
    $\frac S {\sqrt B} @ $ $3~\abi$ &       &       & 4.8  & 2.4  & 2.9  & 14 & 0.7 & 11 \\ \hline
    \end{tabular}%
  \caption{Cut flow table and significance $S/\sqrt B$ for $\zp\to b\bar b$ processes at LHC $14~\tev$. The cross sections $\sigma$ before cuts are for bottom pair (dijet) invariant mass from $2.5-3.5~\tev$. $\epsilon_b$ represents the percentage acceptance of at least one tagged $b$-jet. $\epsilon_{P^{b}_t}$ represents the percentage acceptance also requiring the $p_T$
  of the leading $b$-jet to be greater than $1.2~\tev$. $\sigma_{\rm eff}$ is the cross section after these cuts.}
  \label{tab:LHCbb}%
\end{table}%

For the bottom pair final state
we include both the QCD dijet background and the SM bottom pair irreducible background. We show the cut flow effective acceptance $\epsilon$ and final significance at LHC $14~\tev$ in Table~\ref{tab:LHCbb}. The QCD dijets are required to be in the mass window of $2.5-3.5~\tev$,  with $h_t>500~\gev$ and  leading jet $p_t>200~\gev$ at the parton level. The cross section is $36$ pb, but tight $b$-tagging criteria that have a  $0.1\%$ fake rate from light quark jets can reduce it greatly. Both the signal and irreducible bottom pair background require $b\bar b$ invariant mass in the same window.
The effective invariant mass $m_{\rm eff}$ is the invariant mass of all the jets with $p_t>100~\gev$. After these series of cuts, we will be able to establish three sigma significance for the excess for the benchmark models $Z_{\chi},~Z_{ LR}$ and $Z_{ SSM}$ in the $b\bar b$ final state at LHC $14~\tev$ with $3000~\fbi$.

\subsection{ILC Effects}\label{sec:ILC}

A lepton collider with high luminosity could probe the $\zp$ couplings through their interference with the SM. Here we study the sensitivity  of different observables  to a \zpr\  at the $500~\gev$ and $1~\tev$ ILC. Previous studies include~\cite{DelAguila:1993rw,DelAguila:1995fa,Accomando:1997wt,Weiglein:2004hn,Godfrey:2005pm,Osland:2009dp,Linssen:2012hp,ILCTDR,Battaglia:2012ez}.

\begin{figure}
\subfigure{
\includegraphics[bb=0.0in 0in 7in 6.0in,width=200pt]{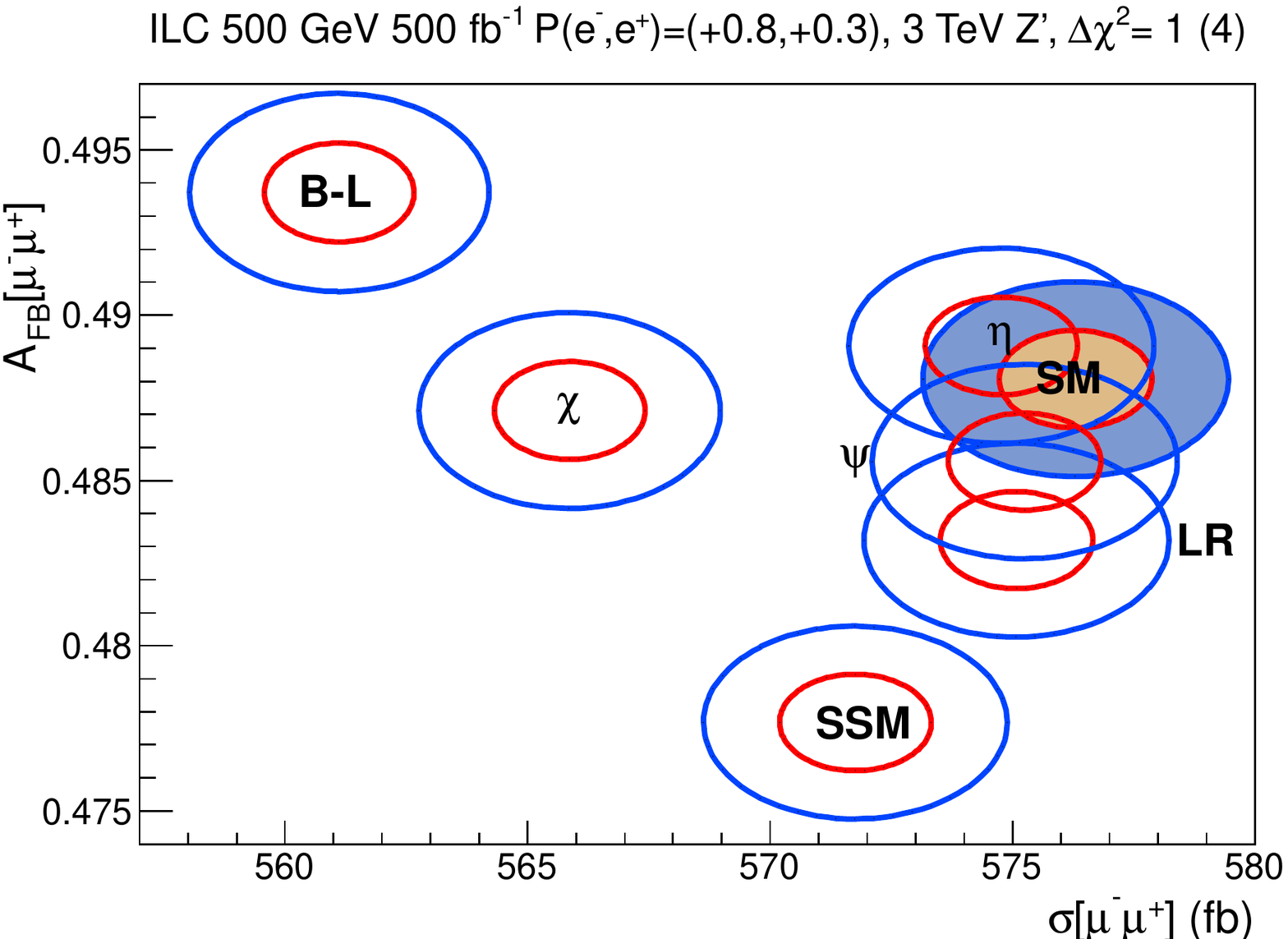}}
\subfigure{
\includegraphics[bb=0.0in 0in 7in 6.0in,width=200pt]{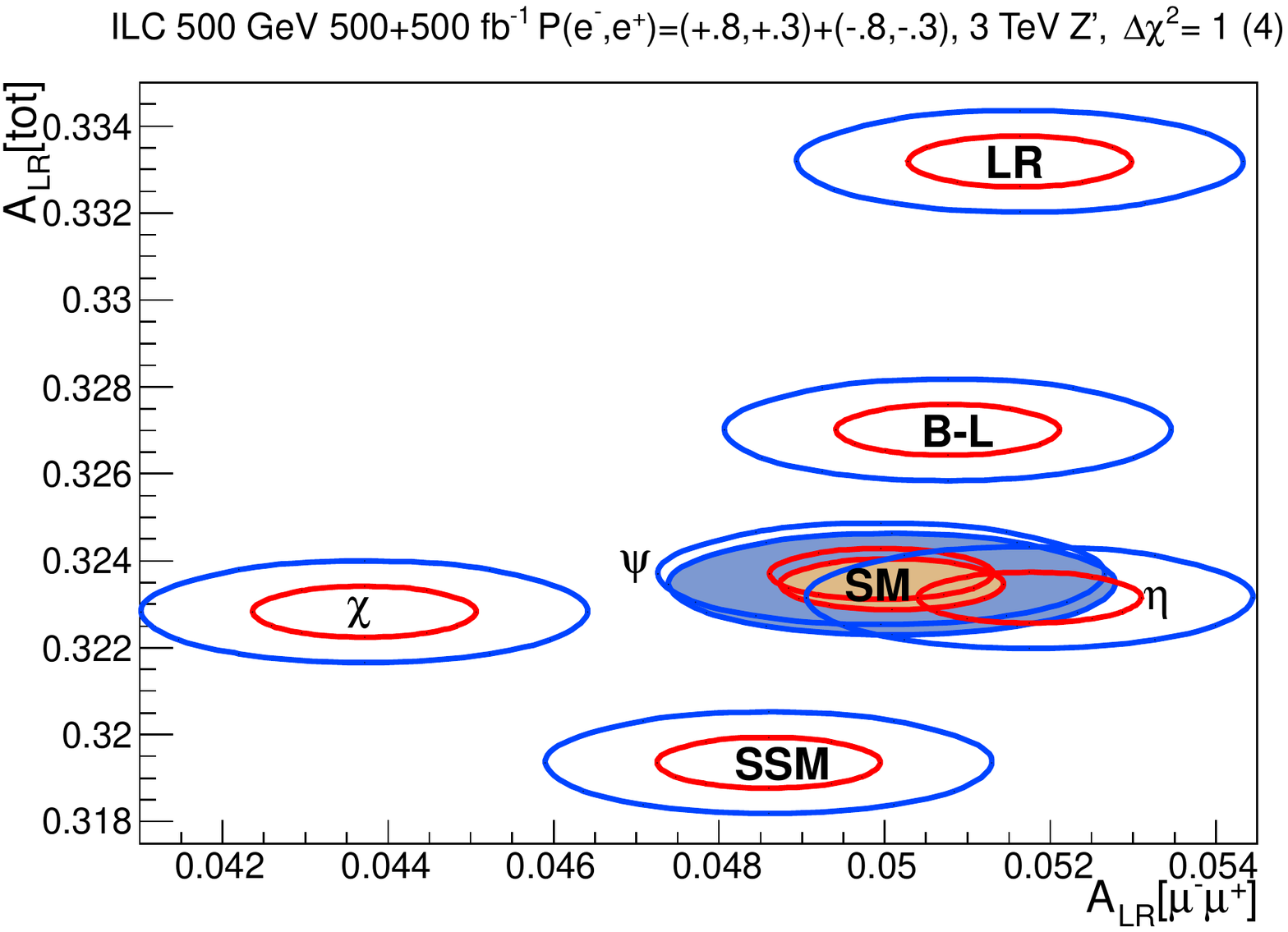}}
\caption[]{The accuracies of ILC observables for a $500~\gev$ ILC. Details of the assumed uncertainties  are discussed in the text. {\it Left panel}: $\Delta\chi^2=1 (4)$ contours (red (blue)) of the simulated $e^-e^+\rightarrow \mu^-\mu^+$ cross sections $\sigma[\mu^-\mu^+]$ and the forward-backward asymmetry  $A_{FB}[\mu^-\mu^+]$ in the dimuon system, with $500~\fbi$ data at fixed beam polarization $P(e^-,e^+)=(+0.8,+0.3)$.
{\it Right panel}: $\Delta\chi^2=1 (4)$ contours (red (blue)) of the simulated polarization (left-right)  asymmetry in the dimuon system $A_{LR}[\mu^-\mu^+]$, and the total polarization asymmetry $A_{LR}[{ tot}]$ (including all of the final states except $e^-e^+$ and $\nu \bar \nu$),  with $500~\fbi$ each for  beam polarizations $P(e^-,e^+)=(+0.8,+0.3)$ and $P(e^-,e^+)=(-0.8,-0.3)$.
}
\label{fig:ILCmmfit}
\end{figure}

We show our results in Figure~\ref{fig:ILCmmfit}. We apply an acceptance of polar angle for the charged leptons in region of $10^\circ<\theta<170^\circ$~\cite{Behnke:2013lya}. We require a minimal $p_T$ of $20~\gev$ for jets. We include a $0.25\%$ polarization uncertainty, $0.2\%$ uncertainties on leptonic observables, and $0.5\%$ uncertainties on hadronic observables~\cite{Baer:2013cma}. Among those uncertainties associated with leptonic and hadronic final states, we assume that $0.14\%$  are correlated and thus will cancel in asymmetry observables. The $\tau$ lepton, bottom quark, and top quark
tagging efficiencies are set at $60\%$, $96\%$~\cite{Baer:2013cma} and $70\%$.

We study the accuracies of the muon forward-backward asymmetry $A_{FB}[\mu^-\mu^+]$ and the cross section $\sigma[\mu^-\mu^+]$ for the dimuon final state\footnote{Dielectron final states also involve $t$-channel exchanges.}, assuming the fixed (normal) beam polarization\footnote{As discussed in Appendix \ref{app:fixed}
we define $\pol >0$  for predominantly left (right)-handed $e^- (e^+)$.} $\pol(e^-,e^+)=(+0.8,+0.3)$, using the formulae in
\refl{ilc1} and \refl{ilc9}.
The muon forward-backward asymmetry in the SM is relatively large, as shown in the {left panel} of Figure~\ref{fig:ILCmmfit}. The difference in cross section is dominantly a summation of  interference terms from different squared helicity amplitudes, and it is possible to have sizable interferences without changing the cross section much. A typical example is the $\zp_{SSM}$, shown in the figure, and similarly the $\zp_{ LR}$. All of the leptonic cross sections are smaller than the SM. This is no longer true for hadronic final states, since $g^e_{L/R}g^{u/d}_{L/R}$ could have either sign. From this figure we can see that $\zp_{\chi}$, $\zp_{ B-L}$ and $\zp_{ SSM}$ are well separated from the SM.

If the  beam polarization can be flipped from normal polarization to the reversed polarization $\pol(e^-,e^+)=(-0.8,-0.3)$, one can determine  the polarization (left-right) asymmetry  $A_{LR}[\mu^-\mu^+]$ for the dimuon channel, defined in \refl{ilc12} and \refl{ilc13}, for which some of the systematics cancel. One can also observe the total polarization asymmetry  $A_{LR}[{ tot}]$
defined in \refl{ilc13a}, for which one does not need to identify the final state (other than removing the dielectron)
and which has higher statistics. However, there are some cancellations between final states. For example, some final states may have positive deviations from the SM while others have negative deviations.
Both $A_{LR}[\mu^-\mu^+]$  and $A_{LR}[{ tot}]$  are shown in the {right panel} of Figure~\ref{fig:ILCmmfit}, assuming 500 fb$^{-1}$ for each polarization\footnote{With the doubled run one would also have such new observables as $\sigma_L + \sigma_R$ in~\refl{ilc12},
$A_{FB}$ in~\refl{ilc9} with reversed polarization, or $A^{FB}_{LR}$ in~\refl{ilc14}. Alternatively, one
could divide a 500 fb$^{-1}$ run into
two 250  fb$^{-1}$  runs with opposite polarizations, in which case the
 outer contours in Figure~\ref{fig:ILCmmfit} would correspond to $\Delta \chi^2=2.6$.}
 For the $A_{LR}[tot]$, we sum  all of the observed final states other than the dielectron\footnote{The major contribution to  $A_{LR}[tot]$ is  from the hadronic final states, since the polarization asymmetry for dileptons is much smaller. One could also consider different final states separately to  gain better statistical sensitivity (but with larger systematic uncertainties).}. $A_{LR}$ has the merit that not only   most of the luminosity uncertainty cancels, but also many systematic uncertainties, such as those associated with tagging efficiencies, acceptances, etc., cancel. Therefore, we only include the polarization  and statistical uncertainties when treating the polarization asymmetries\footnote{Some parametric  uncertainties in the SM parameters don't cancel. We ignore them here as they are expected to improve in the future~\cite{Freitas:2013xga}.}.
 $A_{LR}[\mu^-\mu^+]$ is especially sensitive to $\zp_{\chi}$, while  $A_{LR}[{ tot}]$
is useful for distinguishing $\zp_{ LR}$.

There is some complementarity between the LHC and ILC observations, as can be seen in Figures \ref{fig:LHCmmfit} and \ref{fig:ILCmmfit}. For example, the LHC has limited discrimination
between the LR, B-L, and SSM models, especially from the cleaner $A_{FB}[e^-e^+]$, while
these could be well-separated using the ILC observables.

\section{Case 2: \boldmath{\zpr} Beyond the LHC Reach}\label{sec:heavy}

\begin{figure}
\centering
\includegraphics[bb=0in 0in 6.in 6.0in,width=300pt]{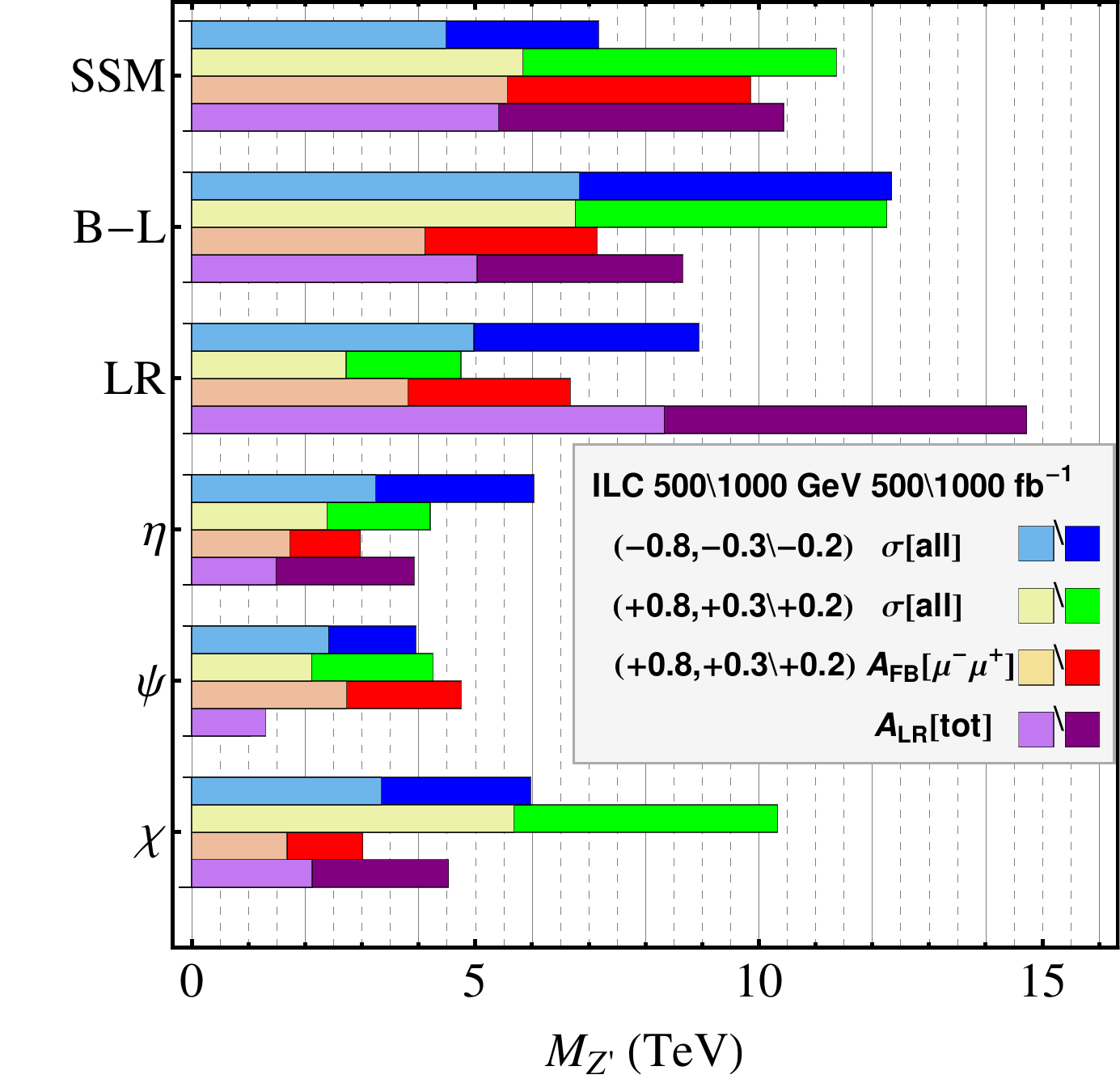}
\caption[]{The exclusion reach of the $500~\gev$ ($1~\tev$) ILC with $500~\fbi$ ($1000~\fbi$) of integrated luminosity for both normal beam polarization $P(e^-,e^+)=(+0.8,+0.3)$ ($P(e^-,e^+)=(+0.8,+0.2)$) (brown (red) and yellow (green)) and reversed beam polarization $P(e^-,e^+)=(-0.8,-0.3)$ ($P(e^-,e^+)=(-0.8,-0.2)$) (cyan (blue)). We show the complementarities between different beam polarizations and observables $\sigma$ including all channels other than the dielectron (cyan (blue) and yellow (green)) and $A_{FB}$ from the dimuon final state (brown (red)). We also show the exclusion reach (magenta (purple)) from $A_{LR}[tot]$ for reversed beam polarizations, with $500+500~\fbi$ and $1000+1000~\fbi$ for the  ILC $500~\gev$ and ILC $1000~\gev$, respectively.
The reaches from  $A_{LR}[tot]$ would be reduced by $\sim$15\% for divided runs of $250+250~\fbi$ and $500+500~\fbi$.
}
\label{fig:reach}
\end{figure}

We show the exclusion (95\% C.L.) reach of the ILC at $500~\gev$ and 1 TeV, including the case that the $\zp$ is beyond the LHC reach,
in Figure~\ref{fig:reach}.  We show the  reach from both the  normal and reversed beam polarizations obtained from the
cross sections for $\mu^- \mu^+, \tau^- \tau^+$, 2 jets (from light quarks), $b \bar b$, and $t \bar t$,
where we combine the $\chi^2$ from each channel after including the appropriate systematic uncertainties. We also show the exclusion reach from the muon forward-backward asymmetry $A_{FB}[\mu^-\mu^+]$ and from  $A_{LR}[tot]$. In the latter case we assume that the
beam polarizations can be reversed and that a full luminosity run is made for each polarization.
The uncertainties included are described in the previous section. We assume the deviations in the cross sections and asymmetries from the SM scale with $\mzp^{-2}$. We conservatively estimate that corrections will reduce the exclusion reach by $\lesssim 2\%$.
There is no single best exclusion observable; for some models like $\zp_{\chi}$ and $\zp_{SSM}$ the normal polarization is better, for others like $\zp_{\psi}$, $\zp_{\eta}$ and $\zp_{LR}$ the reversed beam polarization or the forward-backward asymmetry has a larger reach.
The polarization asymmetries, with a portion of systematic uncertainties cancelled,
is especially stringent for the LR model.

\section{Conclusions}

We study and discuss the \zpr\ discovery and model discrimination potential of the LHC and ILC, using the benchmark models $\zp_\chi$, $\zp_\psi$, $\zp_\eta$, $\zp_{ LR}$, $\zp_{ B-L},$ and $\zp_{ SSM}$. We discuss two  scenarios: (1) a $3~\tev$ $\zp$ that can be resonantly produced at the LHC; (2) a $\zp$ that is too massive to observe as a clear resonance signal.

We discuss the potential of the LHC at $14~\tev$ with integrated luminosity of $300~\fbi$ and $3000~\fbi$ in both leptonic and hadronic final states. The leptonic final states have low background and provide the best sensitivity for discovery. The excellent lepton energy resolution allows them to probe the \zpr\ mass and width. We show  in the left panel of Figure~\ref{fig:LHCmmfit} that for $300~\fbi$ (and $3000~\fbi$), one can reach around $10~\gev$ precision for each at $\sim 1\sigma$. Unfortunately, the width uncertainty is a significant fraction of   the
width itself for typical models with electroweak-scale couplings, limiting the possibility of constraining the absolute
magnitudes of the couplings. The leptonic forward-backward asymmetry, combined with the cross section would have some sensitivity to the chiral couplings, and in particular would allow
discrimination between benchmark models (with minimal width) at a reasonable level. We also discuss  the hadronic \zpr\ modes at the LHC. We study the sensitivity of the bottom pair final state in detail. Although there is a large background from mis-tagged light jets as shown in Table~\ref{tab:LHCbb}, a  $3\sigma$ excess can be achieved for certain benchmark models, such as $\zp_\chi$, $\zp_{ LR}$ and $\zp_{ SSM}$.

For the ILC,  the chiral couplings and \zpr\ mass affect various observables  through the interference  of the \zpr\ with SM contributions. Typical observables include the cross section $\sigma$, forward-backward asymmetry $A_{FB}$ for di-fermion systems with charge identification, and polarization asymmetries $A_{LR}$ for reversed beam polarizations. (Other possibilities include the
polarized forward-backward asymmetry and the final state polarizations in $\tau^+\tau^-$ and $t\bar t$.)
We show the cross section and  forward-backward asymmetry for the
 dimuon system in the left panel of Figure~\ref{fig:ILCmmfit}. It shows good discrimination potential for $\zp_\chi$, $\zp_{B-L}$ and $\zp_{SSM}$ from other models and the SM background.
The polarization asymmetry for the total (except for dielectron) cross section and the dimuon final states are also potentially very useful if the $e^\mp$ polarizations can be reversed, as
 shown in the right panel of Figure~\ref{fig:ILCmmfit}. The asymmetry for the total cross section is
 especially important because it involves high statistics and reduced systematic uncertainties, since the final states do not need to be identified.

For the scenarios in which the $\zp$ cannot be resonantly produced, we study the exclusion reach for the ILC from cross sections, forward-backward asymmetries, and polarization asymmetries. The results are shown in Figure~\ref{fig:reach}, which also shows the complementarity between these observables.

In this preliminary study we have focused on the ability of various observables at the
LHC and ILC to discriminate between several benchmark \zpr\ models with minimal width.
For $\mzp\sim 3$ TeV the LHC should be able to observe a \zpr\ through its
leptonic decays, and obtain a measurement of its mass and width at the 10 GeV level.
Some sensitivity to the chiral couplings (as illustrated by model discrimination)
would be possible at the LHC and especially at the ILC,
and the ILC reach would extend considerably higher as well.

However, there are a very large variety of possible models, including those with much weaker
or stronger couplings than our benchmarks. Ideally, one would like to obtain as much information
as possible in a model-independent way from the LHC, ILC, other colliders, and also
from existing and future precision electroweak experiments.
The inclusion of  additional observables (such as heavy particle final states, additional asymmetries and polarizations, and precision electroweak constraints), a global $\chi^2$ study for model discrimination,   the possibility of model-independent
coupling extractions, and the implications of departing from such assumptions as
family universality are under investigation.

\acknowledgments
We are grateful to Chip Brock and Steve Godfrey for useful discussions.
T.H.~and Z.L.~are supported in part by the U.S.~Department of Energy under Grant No.~DE-FG02-95ER40896, and in part by the PITT PACC.
Z.L.~is also supported in part by the LHC Theory Initiative from the U.S.~National Science Foundation under Grant No.~NSF-PHY-0969510.
L.T.W.~is supported by the NSF under grant PHY-0756966 and the DOE Early Career Award under grant.~de-sc0003930.
This work was supported in part by the National Science Foundation under Grant No.~PHYS-1066293 and the hospitality of the Aspen Center for Physics.

\appendix
\section*{Appendices}
\section{\boldmath{\zpr}  at the LHC}\label{app:LHC}

Here we establish our notation and summarize the basic formalism for the production and decay of a \zpr\ into
a fermion pair at the LHC by the process $p_A\, p_B \ra f \bar f +X$. We define
$s=(p_A + p_B)^2$ and $\hat s  = (p_f + p_{\bar f})^2$, where in our examples we take $s= (14 \text{ TeV})^2$
and  $\hat s  =  (3  \text{ TeV})^2$. $y$ is the $ f \bar f $ rapidity, with $y>0$ along the $\vec p_A$ direction
(i.e., the $f \bar f $ boost direction).
$\theta^* $ is the angle of $f$  in the $ f \bar f $ rest frame, defined\footnote{The $\theta^*$ convention is opposite that  in~\cite{Langacker:1984dc} for $y < 0$, which was motivated by the simultaneous study of $p \bar p$.} with respect to $y$ (i.e., with respect to
$ \vec p_A$ for $y >0$ and $ \vec p_B$ for $y <0$), and $z= \cos\theta^*$. We ignore the transverse momentum $p_T$ of the $f \bar f$ system.

 Let $f_{q_i}(x) [f_{\bar q_i}(x)]$ be the proton PDF of the $i^{th}$ flavor quark [antiquark] $q_i$ [$\bar q_i$],  evaluated at the scale $\mu^2$, which we will take to be $\hat s$.
  The tree-level cross section for Drell-Yan production is then
 \beq
 \frac{d \sigma}{d \hat s\, dy\, dz} = \frac{1}{\hat s} \sum_{i=u,d,c,s,b}
\left[ p_i (\hat s,y)\, \frac{d \sigma ( q_i \bar q_i \ra f \bar f)}{d z} +  \bar p_i \, (\hat s,y) \frac{d \sigma ( \bar q_i  q_i \ra f \bar f)}{d z}\right],
 \eeql{lhc1}
 where\footnote{Higher order QCD $K$ factors $K (\hat s,y)$ can be included in $p_i$ and $\bar p_i$.  We have not implemented the $K$
 factors in the present study. They will potentially increase the sensitivity through an increase in cross section, and may  alter the angular distribution slightly.}
\beq p_i (\hat s,y) \equiv x_A x_B f_{q_i} (x_A) f_{\bar q_i}(x_B), \qquad \bar p_i (\hat s,y) \equiv x_A x_B f_{\bar q_i} (x_A) f_{ q_i}(x_B),
\eeql{lhc2}
with
\beq x_{A,B} \equiv \sqrt{\frac{\hat s}{s}}\ e^{\pm y} .
\eeql{lhc3}
 For family-independent couplings and ignoring quark masses, we can absorb the
 heavier quark PDFs into $p_{u,d}$, i.e., we redefine
$p_u + p_c \ra p_u$ and $p_d + p_s + p_b \ra p_d$, and similarly for $\bar p_{u,d}$,
with $\sum_{i=u,d,c,s,b} \ra \sum_{i=u,d}$. We also define the distribution functions integrated over rapidity
\beq P_i(\hat s, y_1, y_2) = \int_{y_1}^{y_2} p_i (\hat s,y)\, dy, \qquad \overline P_i(\hat s, y_1, y_2) = \int_{y_1}^{y_2} \bar p_i (\hat s,y)\, dy,
\eeql{lhc4}
where $0 \le y_1 < y_2 \le y_{max}$.

The differential cross sections in \refl{lhc1} due to $s$-channel $\gamma, Z$, and \zpr\ are given by
\beq
\frac{d \sigma ( q_i \bar q_i \ra f \bar f)}{d z} = \frac{C_f}{384\pi \hat s}
 \left\{ \left[ G_{LL}^i  + G_{RR}^i \right]  (1+z)^2 + \left[ G_{LR}^i  + G_{RL}^i \right]  (1-z)^2 \right\},
 \eeql{lhc5}
 where $C_f$ is the color factor (1 for leptons and 3 for quarks), and\footnote{$G^i_{ab}$ and the analogous $C^i_{ab}$ defined in \refl{lhc9} should more properly be written as $G^{if}_{ab} $ and $C^{if}_{ab}$, respectively. We usually suppress the dependence on the final state fermion for notational simplicity.}
\beq G^i_{ab} (\hat s) = \left| e^2\, q^i q^f  + \frac{g_a^{1i} g_b^{1f}\,  \hat s }{\hat s -M_Z^2 + i M_Z \Gamma_Z} + \frac{g_a^{2i} g_b^{2f}\,   \hat s }{\hat s -\mzp^2 + i \mzp\gzp}\right|^2
\eeql{lhc6}
for $a,b=L,R$.
The expression for
$\frac{d \sigma ( \bar q_i  q_i \ra f \bar f)}{d z}$ is the same except $(1 \pm z)^2 \ra ( 1 \mp z)^2$.
We have ignored the masses of the initial and final fermions in \refl{lhc5}, which is an adequate approximation
except for the $t$ quark. For our simulations, the top quarks mass is approximately included. The massive top quark will also affect the top charge tagging efficiency through its leptonic decays.
For simplicity, we will evaluate the SM couplings for both the LHC and ILC cases at $M_Z$.

Near the \zpr\ pole it is often adequate to ignore the $\gamma$ and $Z$, in which case
\beq G^i_{ab} (\hat s) \ra \hat s^2 |D(\hat s)|^2  C_{ab}^i ,
\eeql{lhc7}
where
\beq
 |D(\hat s)|^2   = \frac{1}{(\hat s - \mzp^2)^2 + \mzp^2 \gzp^2}
   \eeql{lhc8}
is the Breit-Wigner propagator-squared and
\beq
C_{ab}^i \equiv  |g_a^i|^2 \,  |g_b^f|^2, \qquad a,b=L,R.
\eeql{lhc9}

\subsection{Narrow Width Approximation}\label{app:nwa}
We first consider \zpr\ production, ignoring interference effects, in the narrow width approximation (NWA),
\beq
  |D(\hat s)|^2 \ra \frac{\pi}{\mzp\gzp} \delta(\hat s-\mzp^2).
  \eeql{lhc10}
This is a reasonable first approximation for a multi-TeV scale \zpr\ with electroweak couplings,
for which typically $\gzp/\mzp = \mathcal{O}$(1\%) unless there are important non-SM decay channels.

The cross section is then
\beq
\begin{split}
 \frac{d \sigma}{dy\, dz}& \ra \frac{C_f}{384 \mzp \gzp}
 \sum_{i=u,d} \left\{  \left[ p_i C_N^i + \bar p_i C_F^i \right]  (1+z)^2  + \left[ p_i C_F^i + \bar p_i C_N^i \right]  (1-z)^2  \right\} \\
 &=  \frac{C_f}{384 \mzp \gzp}
 \sum_{i=u,d} \left\{  p_i^+ C^i_+  (1+z^2)  + 2 p_i^- C^i_-  z \right\},
\end{split}
\eeql{lhc11}
  where
\beq p_i^\pm \equiv p_i \pm \bar p_i, \qquad  P_i^\pm \equiv P_i \pm \overline P_i,
\eeql{lhc12}
 and
\beq \begin{split}
C_N^i &\equiv C_{LL}^i + C_{RR}^i , \qquad  C_F^i\equiv C_{LR}^i  + C_{RL}^i  \\
  C^i_\pm &\equiv C_N^i \pm C_F^i = (C_{LL}^i + C_{RR}^i) \pm (C_{LR}^i  + C_{RL}^i).
\end{split}
\eeql{lhc13}

Integrating over angles
(one could include a cut on maximum $|z|$):
\beq\begin{split}
 \frac{d \sigma}{dy} & = \int_{-1}^{+1} \frac{d \sigma}{dy\, dz}  dz
= \frac{C_f}{144 \mzp \gzp} \left\{ p_u^+ C^u_+  +  p_d^+ C^d_+ \right\}  \\
\sigma &=  \left( \int_{y_1}^{y_2} + \int_{-y_2}^{-y_1} \right) \frac{d \sigma}{dy}  dy
= \frac{C_f}{72 \mzp \gzp} \left\{ P_u^+ C^u_+  +  P_d^+ C^d_+ \right\}  .
\end{split}\eeql{lhc14}

These results are sometimes rewritten in terms of the \zpr\ partial widths
\beq \begin{split}
\Gamma(\zpr\ra f \bar f) &= \frac{C_f \mzp}{24\pi} \left( |g_L^f|^2 +  |g_R^f|^2 \right)  \\
\Gamma(\zpr\ra q_i \bar q_i) &= \frac{ \mzp}{8\pi} \left( |g_L^i|^2 +  |g_R^i|^2 \right) ,
\end{split}\eeql{lhc15}
so that
\beq  \frac{d \sigma}{dy} = \frac{4\pi^2}{3 M^3_{\zpr}}\,  [p_u^+ \, \Gamma(\zpr\ra u \bar u) +
p_d^+\, \Gamma(\zpr\ra d \bar d)  ] B(\zpr\ra f \bar f), \eeql{lhc16}
where $B(\zpr\ra f \bar f) \equiv \Gamma(\zpr\ra f \bar f)/\gzp$ is the branching ratio into $f\bar f$.
Similarly,
\beq \sigma
 \equiv \sigma_{\zpr} B(\zpr\ra f \bar f) = \frac{8\pi^2}{3 M^3_{\zpr}}\,  [P_u^+ \, \Gamma(\zpr\ra u \bar u) +
P_d^+\, \Gamma(\zpr\ra d \bar d)  ] B(\zpr\ra f \bar f) \eeql{lhc17}
is the total cross section into $f\bar f$. (We will sometimes denote $\sigma$ by $ \sigma^f $or by $ \sigma[f\bar f]$.)
 Since \gzp\ is not known a priori (except in specific models) one cannot directly constrain
the absolute couplings from $\sigma^f$, although one can obtain ratios of couplings by comparing different final states.
 However, if \gzp\ can be measured independently from the lineshape to a precision of around $25~\gev$ as shown in the left panel of Figure~\ref{fig:ILCmmfit},
then $ \sigma^f\,
\gzp = \sigma_{\zpr} \Gamma(\zpr\ra f \bar f)$ does contains information on the absolute couplings.
Another difficulty is that the cross section for a given $f$ depends on the combination $C^u_+ + C^d_+ (P_d^+/P_u^+)$.
In principle, one could separate $C_{u,d}^+$
by using the rapidity dependence, but in practice there is little sensitivity for $\mzp \gtrsim 3$ TeV.
(Similar statements apply to the rapidity dependence of the angular distribution.) The $u$ and $d$ couplings could, however, be separated if one can observe $b\bar b$ and $t\bar t$ (assuming family-universality).

In addition to \gzp, the total cross sections suffer from PDF,  luminosity,  $K$ factor, and other systematic
uncertainties.
These difficulties are reduced for ratios of rates for different final states, angular distributions, and
final state polarizations.

\subsection{Angular Distribution}\label{app:angular}
Define the forward ($F$) and backward ($B$) cross sections for rapidity $y$ as
\beq F(y) \equiv \int_0^1 \frac{d \sigma}{dy\, dz}  dz, \qquad B(y)  \equiv \int_{-1}^0 \frac{d \sigma}{dy\, dz}  dz. \eeql{lhc18}
Recall that positive $z$ corresponds to $f$ in the direction of the rapidity, so that $F$ and $B$ are symmetric under $y \ra -y$.
It is also useful to define $F$ and $B$ integrated over a range of $|y|$:
\beq  F \equiv  \left( \int_{y_1}^{y_2} + \int_{-y_2}^{-y_1} \right) F(y)\, dy, \qquad B  \equiv  \left( \int_{y_1}^{y_2} + \int_{-y_2}^{-y_1} \right) B(y)\, dy.
\eeql{lhc19}
The forward-backward asymmetries are then
\beq\begin{split}
A_{FB}(y) & \equiv \frac{F(y)-B(y) }{F(y)+B(y)} = \frac{3}{4} \ \frac{p_u^- C^u_-  +  p_d^- C^d_- }{p_u^+ C^u_+  +  p_d^+ C^d_+} \\
A_{FB}& \equiv \frac{F-B}{F+B} = \frac{3}{4}\ \frac{P_u^- C^u_-  +  P_d^- C^d_- }{P_u^+ C^u_+  +  P_d^+ C^d_+},
\end{split}
\eeql{lhc20}
for which the \gzp, luminosity, and some of the PDF uncertainties cancel.
Of course, $A_{FB}(0)=0$ for $pp$ since $p^i_-=0$,  but $A_{FB}(y)$ can
be nonzero for $y\ne 0$~\cite{Langacker:1984dc}. For large positive $y$, for example, the cross section is dominated by $q_i \bar q_i$, with little
dilution from $\bar q_i q_i$, leading to the possibility of a large asymmetry. Of course, the cross section is smaller
at high $y$, so that one should try to optimize the $y_{1,2}$ range.

The forward-backward asymmetry  is equivalent to the charge asymmetry $A_c$ defined by
\beq A_{FB}=A_c\equiv \frac{\sigma( |y_f| > |y_{\bar f}|)-\sigma( |y_f| < |y_{\bar f}|)}{\sigma( |y_f| > |y_{\bar f}|)+\sigma( |y_f|< |y_{\bar f}|)}, \eeql{lh21}
at least in the absence of cuts.

\subsection{Final State Polarization}\label{app:polarization}

One can also consider final state polarizations\footnote{Here we list just the polarizations. In practice, it might be best to consider the actual observables that depend on the polarization, i.e.,  the angular distributions of the $f$  and $\bar f$ decay products.},
defined as
\beq P_f = \frac{\sigma^{f_R}-\sigma^{f_L}}{\sigma^{f_R}+\sigma^{f_L}}, \eeql{lhc22}
where $\sigma^{f_R}$ and $\sigma^{f_L}$ are respectively the rates for producing right and left-helicity $f$.

In addition to \refl{lhc13} it is convenient to define the combinations
\beq\begin{split}
C_L^i & \equiv C_{LL}^i + C_{LR}^i, \qquad C_R^i  \equiv C_{RL}^i + C_{RR}^i \\
\hat C_L^i & \equiv C_{LL}^i + C_{RL}^i, \qquad \hat C_R^i  \equiv C_{LR}^i + C_{RR}^i,
\end{split}
\eeql{lhc23}
and
\beq\begin{split}
C_P^i & \equiv C_L^i-C_R^i = C_{LL}^i - C_{RR}^i + C_{LR}^i - C_{RL}^i  \\
\hat C_P^i & \equiv \hat C_L^i-\hat C_R^i = C_{LL}^i - C_{RR}^i - C_{LR}^i + C_{RL}^i,
\end{split}
\eeql{lhc24}
with
\beq C_L^i+C_R^i =\hat C_L^i+\hat C_R^i = C_N^i + C_F^i =C_+^i. \eeql{lhc25}
Then, ignoring the mass of $f$,
\beq
P_f = - \frac{ \sum_{i=u,d} \left\{  p_i^+ \hat C^i_P  (1+z^2)  + 2 p_i^- C^i_P  z \right\}}{ \sum_{i=u,d} \left\{  p_i^+ C^i_+  (1+z^2)  + 2 p_i^- C^i_-  z \right\}}.\eeql{lhc26}
One can integrate the numerator and denominator separately over the desired ranges of $y$ and $z$. The polarization of $\bar f$ is
opposite to that of $f$ for $m_f\sim 0$.

\subsection{Beyond the Narrow Width Approximation}\label{interference}
Define the combinations $G^i_{N,F}, G^i_\pm, G^i_{L,R}$,  $\hat G^i_{L,R}, G^i_{P}$,  and $\hat G^i_{P}$  of the parameters $G^i_{ab} (\hat s)$ in \refl{lhc6} in analogy with the combinations of $C^i_{ab}$ in \refl{lhc13},
\refl{lhc23}, and \refl{lhc24}.
Then
 \beq\begin{split}
 \frac{d \sigma}{d \hat s\, dy\, dz}
 &=
 \frac{C_f}{384 \pi \hat s^2}
 \sum_{i=u,d} \left\{  \left[ p_i G_N^i + \bar p_i G_F^i \right]  (1+z)^2  + \left[ p_i G_F^i + \bar p_i G_N^i \right]  (1-z)^2  \right\} \\
 &=   \frac{C_f}{384 \pi \hat s^2}
 \sum_{i=u,d} \left\{  p_i^+ G^i_+  (1+z^2)  + 2 p_i^- G^i_-  z \right\},
\end{split}     \eeql{lhc27}
Other relevant observables (for $m_f=0$) are then
\beq \begin{split}
 \frac{d \sigma}{d \hat s\,  dz}&= \frac{C_f}{192 \pi \hat s^2}
 \sum_{i=u,d} \left\{  P_i^+ G^i_+  (1+z^2)  + 2P_i^- G^i_-  z \right\} \\
  \frac{d \sigma}{d \hat s\, dy} &=  \frac{C_f}{144 \pi \hat s^2}      \left\{ p_u^+ G^u_+  +  p_d^+ G^d_+ \right\}  \\
 \frac{d \sigma}{d  \sqrt{\hat s}}&= 2  \sqrt{\hat s}\,   \frac{d \sigma}{d \hat s}   = \frac{C_f}{36 \pi \hat s^{3/2}}      \left\{ P_u^+ G^u_+  +  P_d^+ G^d_+ \right\}  \\
 A_{FB}(\hat s, y)& = \frac{3}{4}\ \frac{p_u^- G^u_-  +  p_d^- G^d_- }{p_u^+ G^u_+  +  p_d^+ G^d_+}  \\
P_f &= - \frac{ \sum_{i=u,d} \left\{  p_i^+ \hat G^i_P  (1+z^2)  + 2 p_i^- G^i_P  z \right\}}{ \sum_{i=u,d} \left\{  p_i^+ G^i_+  (1+z^2)  + 2 p_i^- G^i_-  z \right\}}.
\end{split}\eeql{lhc28}
One can separately integrate the numerator and denominator of $ A_{FB}$ over the desired ranges of $\hat s$ and $y$ to obtain the integrated asymmetry. Similarly,  the numerator and denominator of
$P_f$ can be separately integrated over  the desired ranges of $\hat s$, $y$, and $z$. The polarization of $\bar f$ is
opposite to that of $f$ for $m_f\sim 0$.

\subsection{Finite Mass Corrections}\label{app:finitemt}

\section{\boldmath{\zpr}  at the ILC}\label{app:ILC}
We now consider $e^- e^+ \ra f \bar f$ at CM energy $\sqrt{s}$. The final fermion $f$ can
be $\mu, \tau, b, t$ or possibly $c, s,$ or unidentified quark. (We do not consider $f=e$ because that involves
$t$ channel exchange as well as $s$ channel.)
Define
\beq
 G^e_{ab} ( s) = \left| e^2\, q^e q^f  + \frac{g_a^{1e} g_b^{1f}\,   s }{ s -M_Z^2 + i M_Z \Gamma_Z} + \frac{g_a^{2e} g_b^{2f}\,   s }{ s -\mzp^2 + i \mzp\gzp}\right|^2,
\eeql{ilc1}
in analogy to \refl{lhc6}. We assume  $M_Z \ll \sqrt{s} \ll \mzp$, so we can  ignore $\Gamma_Z$ and $\gzp$.

\subsection{No Polarization}\label{app;unpolarized}
In the absence of polarization for the $e^\mp$ the observables are
\beq
\begin{split}
\frac{d \sigma (s)}{d z} &= \frac{C_f}{128\pi  s}
 \left\{ \left[ G_{LL}^e  + G_{RR}^e \right]  (1+z)^2 + \left[ G_{LR}^e + G_{RL}^e \right]  (1-z)^2 \right] \\
 &= \frac{C_f}{128\pi  s}  \left\{  G^e_+  (1+z^2)  + 2  G^e_-  z \right\}  \\
 \sigma (s) &= \frac{C_f}{48\pi  s}   G^e_+  \\
 A_{FB} (s)& = \frac{3}{4}\ \frac{G^e_-}{G^e_+}  \\
 P_f &= - \frac{\hat G^e_P  (1+z^2)  + 2  G^e_P  z}{G^e_+  (1+z^2)  + 2 G^e_-  z},
\end{split}
\eeql{ilc2}
where the various $G^e_{ab}$ combinations are defined in analogy to  the combinations of $C^i_{ab}$  in  \refl{lhc13},
\refl{lhc23}, and \refl{lhc24}. As usual, one can integrate numerator and denominator of $P_f$ over $z$ to obtain  the average polarization
$P_f = - \hat G^e_P/G^e_+$.

Although we are mainly concerned with the regime  $M_Z \ll \sqrt{s} \ll \mzp$ it is nevertheless useful to display
the asymmetries and polarizations at the $Z$ or \zpr\ pole, ignoring interferences. For $s=M_Z^2$,
\beq  \begin{split}
A_{FB} (M_Z^2)& \ra \frac{3}{4}\ A_e^1 A_f^1  \\
  P_f(M_Z^2)& \ra  - \frac{A_f  (1+z^2)  + 2  A_e  z}{ (1+z^2)  + 2 A_e A_f  z} \ra - A^1_f
\end{split}\eeql{ilc3}
with
\beq  A_f^1 \equiv  \frac{(g_L^{1f})^{2} - (g_R^{1f})^{2}  }{(g_L^{1f})^{2} + (g_R^{1f})^{2} }
= \frac{2\, g_V^{1f}\, g_A^{1f}}{(g_V^{1f})^{2} + (g_A^{1f})^{2}}. \eeql{ilc4}
The second form for $P_f$ is the average polarization.
Similar expressions hold at the \zpr\ pole, with $A^1_f \ra A^2_f$ and $g_a^{1f}\ra g_a^{2f}$.

\subsection{Fixed Initial State Polarization}\label{app:fixed}
For $V$ and $A$ interactions (and ignoring $m_e$), only the combinations $e^-_Le^+_R$ and $e^-_Re^+_L$
contribute  yield nonzero amplitudes (unlike, $S, P,$ and $T$, which are sensitive to $e^-_Le^+_L$ and $e^-_Re^+_R$).
We define the initial state polarizations
\beq  \pol^- = \eta^-_L - \eta^-_R, \qquad \pol ^+ = \eta^+_R - \eta^+_L, \eeql{ilc5}
where $\eta^-_L$ and $\eta^-_R = 1 - \eta^-_L$ are respectively the fractions of $L$ and $R$-helicity
$e^-$, and similarly for $e^+$. Note that (neglecting $m_e$) $\pol ^-=\pol ^+ \sim 1$ for $e^\mp$ produced in
weak charge current processes. Also note that the definition of $\pol^-$ is conventional for $e^- e^+$, though it is opposite in sign from usual polarization definitions. Some useful relations are
\beq   \eta^-_{L,R}= \frac{1\pm \pol^-}{2}, \qquad  \eta^+_{L,R}= \frac{1\mp \pol^+}{2} \eeql{ilc6}
\beq  \frac{ \eta^-_L \eta^+_R}{ \eta^-_L \eta^+_R+ \eta^-_R \eta^+_L }=\frac{1+ \pol_{eff}}{2}, \qquad
\frac{\eta^-_R \eta^+_L }{ \eta^-_L \eta^+_R+ \eta^-_R \eta^+_L }=\frac{1- \pol_{eff}}{2}, \eeql{ilc7}
where the effective polarization is defined as
\beq   \pol _{eff} \equiv \frac{\pol ^- + \pol ^+}{1+ \pol ^- \pol ^+} = \frac{ \eta^-_L \eta^+_R - \eta^-_R \eta^+_L }{\eta^-_L \eta^+_R + \eta^-_R \eta^+_L}.\eeql{ilc8}
For example, $\pol ^-=0.80$ and $\pol ^+ = 0.30$ yields $\pol _{eff} \sim 0.89$, while
$(\pol ^-, \pol ^+) = (0.80, 0.60) \Rightarrow \pol _{eff} \sim 0.95.$

The relevant observables for fixed polarizations are
\beq  \begin{split}
\frac{d \sigma}{d z} &= \frac{C_f}{32\pi  s}
 \left\{ \left[  \eta^-_L \eta^+_R\, G_{LL}^e  +\eta^-_R \eta^+_L \, G_{RR}^e \right]  (1+z)^2 \right. \\
& \qquad\qquad  \left. + \left[  \eta^-_L \eta^+_R\, G_{LR}^e +\eta^-_R \eta^+_L\,  G_{RL}^e \right]  (1-z)^2 \right\} \\
 \sigma &= \frac{C_f}{12\pi  s}   \left[  \eta^-_L \eta^+_R\, G_{L}^e  +\eta^-_R \eta^+_L \, G_{R}^e \right]   \\
 A_{FB} & = \frac{3}{4}\ \frac{G^e_-+ \pol _{eff}\, \hat G_P^e}{G^e_+ +  \pol _{eff}\,  G_P^e} \ra
 \frac{3}{4}\ A_f \frac{A_e +  \pol _{eff}}{1+ \pol _{eff}\, A_e}  \\
 P_f &= - \frac{\bigl[ \hat G^e_P +  \pol _{eff}\,  G_-^e \bigr]  (1+z^2)  + 2 \bigl[ G^e_P+  \pol _{eff}\,  G_+^e  \bigr]  z}{\bigl[  G^e_+ +  \pol _{eff}\,  G_P^e \bigr]  (1+z^2)  + 2  \bigl[ G^e_-+  \pol _{eff}\,  \hat G_P^e  \bigr]  z}  \\
 &\quad  \ra
  - \frac{\left[ A_f +  \pol _{eff}\,  A_e A_f  \right]  (1+z^2)  + 2 \left[ A_e +  \pol _{eff} \right]  z}{\left[  1 +  \pol _{eff}\,  A_e \right]  (1+z^2)  + 2  \left[ A_e A_f  + \pol _{eff}\,  A_f  \right]  z}  \\
  &\quad \ra - A_f.
\end{split}
 \eeql{ilc9}
The second forms for $ A_{FB}$ and $P_f$ are valid at the $Z$ or \zpr\ pole (the superscript 1 or 2 on $A_{e,f}$ is implied). The third form for $P_f$ is obtained by integrating the numerator and denominator  over $z$.

\subsection{Polarization Asymmetries}\label{app:polasym}
Denote the cross sections for the polarizations $\pol^\mp$ defined in the previous section by $\sigma_L$,
and let $\sigma_R$ represent the cross section for reversed polarizations
\beq  \eta_{L,R}^- \ra  \bar \eta_{L,R}^- , \qquad  \eta_{L,R}^+\ra  \bar \eta_{L,R}^+, \eeql{ilc10}
with
\beq   \bar \eta_{L,R}^-= \eta_{R,L}^-, \qquad \bar \eta_{L,R}^+= \eta_{R,L}^+,\eeql{ilc11}
so that $\pol^\mp \ra - \pol^\mp$. For example,
\beq  \begin{split}
\sigma_L &=  \frac{C_f}{12\pi  s}  \left\{  \eta^-_L \eta^+_R G^e_L + \eta^-_R \eta^+_L G_R^e \right\}  \\
\sigma_R &=  \frac{C_f}{12\pi  s}  \left\{\bar  \eta^-_L\bar \eta^+_R G^e_L + \bar \eta^-_R\bar \eta^+_L G_R^e \right\}
= \frac{C_f}{12\pi  s}  \left\{  \eta^-_R \eta^+_L G^e_L + \eta^-_L \eta^+_R G_R^e \right\} .
\end{split}
\eeql{ilc12}
For $\pol ^\pm = 0$ these both reduce to the unpolarized cross section~$\sigma$.

The polarization (left-right) asymmetry is defined as
\beq  A_{LR} \equiv \frac{\sigma_L - \sigma_R}{\sigma_L +\sigma_R} = \pol_{eff} \frac{G^e_P}{G^e_+} \ra  \pol_{eff} A_e,\eeql{ilc13}
where the second form is valid on the $Z$ or \zpr\ pole. (At the pole, $A_{LR}$ is independent of the final state,
allowing the determination of $A_e$ from the total cross section polarization asymmetry.)
It is also useful to define the total polarization asymmetry
\beq  A_{LR}^{tot} \equiv \frac{\sigma_L^{tot}  - \sigma_R^{tot} }{\sigma_L^{tot}  +\sigma_R^{tot} } = \pol_{eff} \frac{\sum_f C_f G^{ef}_P}{\sum_f C_f G^{ef}_+} \ra  \pol_{eff} A_e ,\eeql{ilc13a}
where we have added the superscript to emphasize the final state $f$. The sum can be taken over $f=\mu, \tau, u, d, c, s, b,$ and $t$ (if one ignores $m_t$). $ A_{LR}^{tot}$ is convenient in that one does not have to identify the final state (other than removing $f=e$, which also has $t$-channel contributions) and because one therefore has much higher statistics. However, the asymmetries between different final states may partially cancel away from the poles.

Assuming that the $e^-$ and $e^+$ polarizations can each be turned off or reversed without affecting the magnitudes,  one could in principle determine $G^e_P/G^e_+$
(or the analogous quantity in \refl{ilc13a}), $ \pol_{eff}$, $\pol^-$,
and $\pol^+$ experimentally by measuring the asymmetries obtained by reversing the polarizations $(\pol^-,\pol^+ )$, $(\pol^-,0)$, and $(0,\pol^+ )$
(the Blondel scheme~\cite{Blondel:1987wr}).

Another useful observable is the left-right forward-backward asymmetry
\beq A^{FB}_{LR} \equiv \frac{F_L-B_L -F_R +B_R}{F_L+B_L +F_R +B_R}= \frac{3}{4}\ \pol_{eff} \frac{\hat G_P^e}{G_+^e} \ra \frac{3}{4}\ \pol_{eff} A_f, \eeql{ilc14}
where
\beq F_{L,R} \equiv \int_0^1 \frac{ d \sigma_{L,R}}{d z} d z, \qquad B_{L,R} \equiv \int_{-1}^0 \frac{ d \sigma_{L,R}}{d z} d z .\eeql{ilc15}

  One can also define the final state polarization left-right asymmetry:
\beq \begin{split}
 P_f^{LR}& \equiv \frac{\sigma_L^{f_R} -\sigma_L^{f_L} -\sigma_R^{f_R} +\sigma_R^{f_L}  }{\sigma_L^{f_R} +\sigma_L^{f_L} +\sigma_R^{f_R} +\sigma_R^{f_L}  }
= -\pol_{eff}\, \frac{G^e_- (1+z^2) + 2 G^e_+ z}{G^e_+ (1+z^2) + 2 G^e_- z}  \\
&\quad \ra -\pol_{eff}\, \frac{A_e A_f(1+z^2) + 2  z}{ (1+z^2) + 2 A_e A_f z} \ra - \pol_{eff}\, A_e A_f.
\end{split}
\eeql{ilc16}

\bibliographystyle{JHEP}
\bibliography{zprime_snowmass2013}

\end{document}